%% file: paper1663.tex
\documentclass[runningheads]{llncs}
\usepackage[T1]{fontenc}
\usepackage{graphicx}
\usepackage{hyperref}
\usepackage{color}
\usepackage{amsmath}
\usepackage{booktabs}
\usepackage{multirow}

\usepackage{tablefootnote}
\usepackage{adjustbox} 
\usepackage{tabularx}
\newcolumntype{Y}{>{\centering\arraybackslash}X}
\usepackage{bbold} 
\usepackage{arydshln} 
\usepackage{caption} 

\begin{document}

\title{MultiTalent: A Multi-Dataset Approach to Medical Image Segmentation}
%
%


\author{Constantin Ulrich\inst{1,4,5} \and
Fabian Isensee\inst{1,2} \and
Tassilo Wald\inst{1,2} \and
Maximilian Zenk\inst{1,5} \and
Michael Baumgartner\inst{1,2,6} \and
Klaus H. Maier-Hein\inst{1,3}}
%

\authorrunning{C. Ulrich et al.}
%
\institute{Division of Medical Image Computing,\\German Cancer Research Center (DKFZ), Heidelberg, Germany\\
\and Helmholtz Imaging, DKFZ, Heidelberg, Germany\\
\and Pattern Analysis and Learning Group, Department of Radiation Oncology, Heidelberg University Hospital, Heidelberg, Germany \\
\and National Center for Tumor Diseases (NCT), NCT Heidelberg, A partnership between DKFZ and University Medical Center Heidelberg\\
\and Medical Faculty Heidelberg, University of Heidelberg, Heidelberg, Germany
\and Faculty of Mathematics and Computer Science, Heidelberg University, Germany
\email{constantin.ulrich@dkfz-heidelberg.de}}
\maketitle 
\begin{abstract}
The medical imaging community generates a wealth of data-sets, many of which are openly accessible and annotated for specific diseases and tasks such as multi-organ or lesion segmentation. 
Current practices continue to limit model training and supervised pre-training to one or a few similar datasets, neglecting the synergistic potential of other available annotated data. 
We propose MultiTalent, a method that leverages multiple CT datasets with diverse and conflicting class definitions to train a single model for a comprehensive structure segmentation. Our results demonstrate improved segmentation performance compared to previous related approaches, systematically, also compared to single-dataset training using state-of-the-art methods, especially for lesion segmentation and other challenging structures. We show that MultiTalent also represents a powerful foundation model that offers a superior pre-training for various segmentation tasks compared to commonly used supervised or unsupervised pre-training baselines. 
Our findings offer a new direction for the medical imaging community to effectively utilize the wealth of available data for improved segmentation performance. The code and model weights will be published here: \url{https://github.com/MIC-DKFZ/MultiTalent}.

\end{abstract}
\section{Introduction}

\begin{figure}[t]
\centering
\includegraphics[width=1\textwidth]{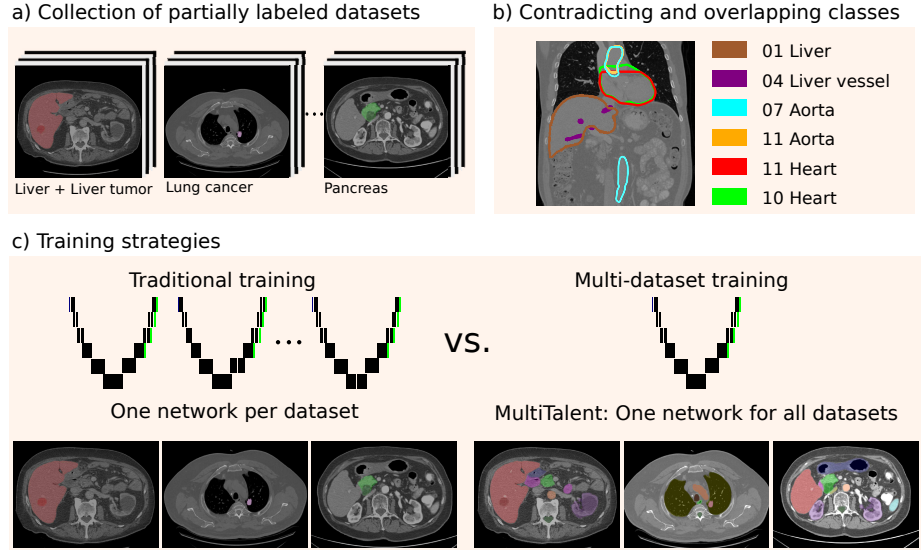}
\caption{(a) Usually only a few classes are annotated in publicly available datasets. b) Different groundtruth label properties can generate contradicting class predictions. For example, the heart annotation of dataset 11 differs from the heart annotation of dataset 10, which causes the aorta of dataset 11 to overlap with the heart of dataset 10. In contrast to dataset 11, in dataset 7 the aorta is also annotated in the lower abdomen. c) Instead of training one network for each dataset, we introduce a method to train one network with all datasets, while retaining dataset-specific annotation protocols.} \label{overview_figure}
\end{figure}

The success of deep neural networks heavily relies on the availability of large and diverse annotated datasets across a range of computer vision tasks. To learn a strong data representation for robust and performant medical image segmentation, huge datasets with either many thousands of annotated data structures or less specific self-supervised pretraining objectives with unlabeled data are needed \cite{ZHOU2021101840,swinunetr}. 
The annotation of 3D medical images is a difficult and laborious task. Thus, depending on the task, only a bare minimum of images and target structures is usually annotated. This results in a situation where a zoo of partially labeled datasets is available to the community. 
Recent efforts have resulted in a large dataset of >1000 CT images with >100 annotated classes each, thus providing more than 100,000 manual annotations which can be used for pre-training \cite{totalseg}. Focusing  on such a dataset prevents leveraging the potentially precious additional information of the above mentioned other datasets that are only partially annotated. Integrating information across different datasets potentially yields a higher variety in image acquisition protocols, more anatomical target structures or details about them as well as information on different kinds of pathologies. 
Consequently, recent advances in the field allowed utilizing partially labeled datasets to train one integrated model \cite{partially_survey}. Early approaches handled annotations that are present in one dataset but missing in another by considering them as background \cite{roulet_joint_2019,fang_multi-organ_2020} and penalizing overlapping predictions by taking advantage of the fact that organs are mutually exclusive \cite{shi_marginal_2020,fidon_label-set_2021}. Some other methods only predicted one structure of interest for each forward pass by incorporating the class information at different stages of the network \cite{dmitriev_learning_2019,zhang_dodnet_2020,clip_medicalseg}. Chen et al. trained one network with a shared encoder and separate decoders for each dataset to generate a generalized encoder for transfer learning \cite{chen_med3d_2019}. However, most approaches are primarily geared towards multi-organ segmentation as they do not support overlapping target structures, like vessels or cancer classes within an organ \cite{feng2021mskd,filbrandt_learning_2021,contextaware,teacher_student}.
So far, all previous methods do not convincingly leverage cross-dataset synergies. 
As Liu et al. pointed out, one common caveat is that many methods force the resulting model to average between distinct annotation protocol characteristics \cite{clip_medicalseg} by combining labels from different datasets for the same target structure (visualized in Figure 1 b)). Hence, they all fail to reach segmentation performance on par with cutting-edge single dataset segmentation methods. 
To this end, we introduce MultiTalent (MULTI daTAset LEarNing and pre-Training), a new, flexible, multi-dataset training method: 1) MultiTalent can handle classes that are absent in one dataset but annotated in another during training. 2) It retains different annotation protocol characteristics for the same target
structure and 3) allows for overlapping target structures with
different level of detail such as liver, liver vessel and liver tumor. Overall, MultiTalent can include all kinds of new datasets irrespective of their annotated target structures. \\ MultiTalent can be used in two scenarios: First, in a combined multi-dataset (MD) training to generate one foundation segmentation model that is able to predict all classes that are present in any of the utilized partially annotated datasets, and second, for pre-training to leverage the learned representation of this foundation model for a new task. In experiments with a large collection of abdominal CT datasets, the proposed model outperformed state-of-the-art segmentation networks that were trained on each dataset individually as well as all previous methods that incorporated multiple datasets for training. Interestingly, the benefits of MultiTalent are particularly notable for more difficult classes and pathologies. In comparison to an ensemble of single dataset solutions, MultiTalent comes with shorter training and inference times.\\
Additionally, at the example of three challenging datasets, we demonstrate that fine-tuning MultiTalent yields higher segmentation performance than training from scratch or initializing the model parameters using unsupervised pre-training strategies \cite{swinunetr,ZHOU2021101840}. It also surpasses supervised pretrained and fine-tuned state-of-the art models on most tasks, despite requiring orders of magnitude less annotations during pre-training.

\section{Methods}
We introduce MultiTalent, a multi dataset learning and pre-training method, to train a foundation medical image segmentation model. It comes with a novel dataset and class adaptive loss function. The proposed network architecture enables the preservation of all label properties, learning overlapping classes and the simultaneous prediction of all classes. Furthermore, we introduce a training schedule and dataset preprocessing which balances varying dataset size and class characteristics. 

\subsection{Problem definition}
We begin with a dataset collection of $K$ datasets $D^{(k)}, k \in [1, K]$, with $N^{(k)}$ image and label pairs $D^{(k)} = \{(x,y)^{(k)}_1,..., (x,y)^{(k)}_{N^{(k)}}\}$. In these datasets, every image voxel $x^{(k)}_i, i \in [1,I]$, is assigned to one class $c \in C^{(k)}$, where $C^{(k)} \subseteq C$ is the label set associated to dataset $D^{(k)}$. Even if classes from different datasets refer to the same target structure we consider them as unique, since the exact annotation protocols and labeling characteristics of the annotations are unknown and can vary between datasets: $C^{(k)} \cap C^{(j)} = \emptyset, \forall k \neq j$.
This implies that the network must be capable of predicting multiple classes for one voxel to account for the inconsistent class definitions. 

\subsection{MultiTalent}
\subsubsection{Network modifications}
We employ three different network architectures, which are further described below, to demonstrate that our approach is applicable to any network topology. To solve the label contradiction problem we decouple the segmentation outputs for each class by applying a Sigmoid activation function instead of the commonly used Softmax activation function across the dataset. The network shares the same backbone parameters $\Theta$ but it has independent segmentation head parameters $\Theta_c$ for each class. The Sigmoid probabilities for each class are defined as $\hat{y}_c=f(x,\Theta, \Theta_c)$. This modification allows the network to assign multiple classes to one pixel and thus enables overlapping classes and the conservation of all label properties from each dataset. Consequently, the segmentation of each class can be thought of as a binary segmentation task.

\subsubsection{Dataset and class adaptive loss function}
\label{loss_function}
Based on the well established combination of a Cross-entropy and Dice loss for single dataset medical image segmentation, we employ the binary Binary Cross-entropy loss (BCE) and a modified Dice loss for each class over all $B, b\in [1,B]$, images in a batch: 

\begin{equation}
     L_c = \frac{1}{I}\sum_{b,i} BCE(\hat{y}^{(k)}_{i,b,c}, \; y^{(k)}_{i,b,c})  -\frac{2\sum_{b,i}\hat{y}^{(k)}_{i,b,c} \; y_{{i,b,c}}^{(k)}} {\sum_{b,i}\hat{y}^{(k)}_{i,b,c} +\sum_{b,i} y_{i,b,c}^{(k)}}
\end{equation}
While the regular dice loss is calculated for each image within a batch, we calculate the dice loss jointly for all images of the input batch. This regularizes the loss if only a few voxels of one class are present in one image and a larger area is present in another image of the same batch. Thus, an inaccurate prediction of a few pixels in the first image has a limited effect on the loss. In the following, we unite the sum over the image voxels $i$ and the batch $b$ to $\sum_{z}$.
We modify the loss function to be calculated only for classes that were annotated in the corresponding partially labeled dataset \cite{roulet_joint_2019,fang_multi-organ_2020}, in the following indicated by $\mathbb{1}_c^{(k)}$, where $\mathbb{1}_c^{(k)} = 1$ if $c \in C^{(k)}$ and 0 otherwise. Instead of averaging, we add up the loss over the classes. Hence, the loss signal for each class prediction does not depend on the number of other classes within the batch. This compensates for the varying number of annotated classes in each dataset. Otherwise, the magnitude of the loss e.g for the liver head from D1 (2 classes) would be much larger as for D7 (13 classes). Gradient clipping captures any potential instability that might arise from a higher loss magnitude:
\begin{equation}
    L = \sum_{c} \Big( \mathbb{1}_c^{(k)} \frac{1}{I}\sum_{z} BCE(\hat{y}^{(k)}_{z,c}, \;y^{(k)}_{z,c})  -\frac{2\sum_{z}\mathbb{1}_c^{(k)} \; \hat{y}^{(k)}_{z,c} \; y_{z,c}^{(k)}} {\sum_{z}\mathbb{1}_c^{(k)} \; \hat{y}^{(k)}_{z,c} +\sum_{z}\mathbb{1}_c^{(k)} \; y_{z,c}^{(k)}} \Big)
\end{equation}


\subsubsection{Network architectures}
To demonstrate the general applicability of this approach, we applied it to three segmentation networks. We employed a 3D U-Net \cite{ronneberger_u-net_2015}, an extension with additional residual blocks in the encoder (Resenc U-Net), that demonstrated highly competitive results in previous medical image segmentation challenges \cite{extendingnnunet,fabi_kits} and a recently proposed transformer based architecture (SwinUNETR \cite{swinunetr}).
We implemented our approach in the nnU-Net framework \cite{isensee_nnu-net_2021}. However, the automatic pipeline configuration from nnU-net was not used in favor of a manually defined configuration that aims to reflect the peculiarities of each of the datasets, irrespective of the number of training cases they contain. We manually selected a patch size of $[96, 192,192]$ and image spacing of 1mm in plane and 1.5mm for the axial slice thickness, which nnU-Net used to automatically create the two CNN network topologies. For the SwinUNETR, we adopted the default network topology.

\subsubsection{Multi-dataset training setup}
\label{Training_strategy}
We trained MultiTalent with 13 public abdominal CT datasets with a total of 1477 3D images, including 47 classes (Multi-dataset (MD) collection) \cite{Antonelli2022,BTCV,BTCV2,structseg,lambert2019segthor,roth2015deeporgan,clark_cancer_2013,NHpancreas,heller2020kits19}. Detailed information about the datasets, can be found in the appendix in Table 3 and Figure 3, including the corresponding annotated classes. 
We increased the batch size to 4 and the number of training epochs to 2000 to account for the high number of training images. To compensate for the varying number of training images in each dataset, we choose a sampling probability per case that is inversely proportional to $\sqrt{n}$, where $n$ is the number of training cases in the corresponding source dataset. Apart from that, we have adopted all established design choices from nnU-Net to ensure reproducibility and comparability. 

\subsubsection{Transfer learning setup} We used the BTCV (small multi organ dataset \cite{BTCV}), AMOS (large multi organ dataset \cite{ji2022amos}) and KiTS19 (pathology dataset \cite{heller2020kits19}) datasets to evaluate the generalizability of the MultiTalent features in a pre-training and fine tuning setting. Naturally, the target datasets were excluded from the repective pre-training. Fine tuning was performed with identical configuration as the source training, except for the batch size which was set to 2. We followed the fine-tuning schedule proposed by Kumar et al. \cite{transferschedule}. First, the segmentation heads were warmed up over 10 epochs with linearly increasing learning rate, followed by a whole-network warum-up over 50 epochs. Finally, we continued with the standard nnU-Net training schedule.

\subsection{Baselines}
As a baseline for the MultiTalent, we applied the 3D U-Net generated by the nnU-Net without manual intervention to each dataset individually. Furthermore, we trained a 3D U-Net, a Resenc U-Net and a SwinUNETR with the same network topology, patch and batch size as our MultiTalent for each dataset. All baseline networks were also implemented within the nnU-Net framework and follow the default training procedure. Additionally, we compare MultiTalent with related work on the public BTCV leaderboard in \autoref{BTCV_leaderboard}.\\
Furthermore, the utility of features generated by MultiTalent is compared to supervised and unsupervised pre-training baselines. As supervised baseline, we used the weights resulting from training the three model architectures on the TotalSegmentator dataset, which consists of 1204 images and 104 classes \cite{totalseg}, resulting in more than $10^5$ annotated target structures. In contrast, MultiTalent is only trained with about 3600 annotations. We used the same patch size, image spacing, batch size and number of epochs as for the MultiTalent training. As unsupervised baseline for the CNNs, we pre-trained the networks on the Multi-dataset collection based on the work of Zhou at al. (Model Genesis \cite{ZHOU2021101840}).
Finally, for the SwinUNETR architecture, we compared the utility of the weights from our MultiTalent with the ones provided by Tan et al. who performed self-supervised pre-training on 5050 CT images. This necessitated the use of the original (org.) implementation of SwinUNETR because the recommended settings for fine tuning were used. This should serve as additional external validation of our model. To ensure fair comparability, we did not scale up any models. Despite using gradient checkpointing, the SwinUNETR models requires roughly 30 GB of GPU memory, compared to less than 17 GB for the CNNs.

\section{Results}
\begin{figure}[t]
\centering
\includegraphics[width=1\textwidth]{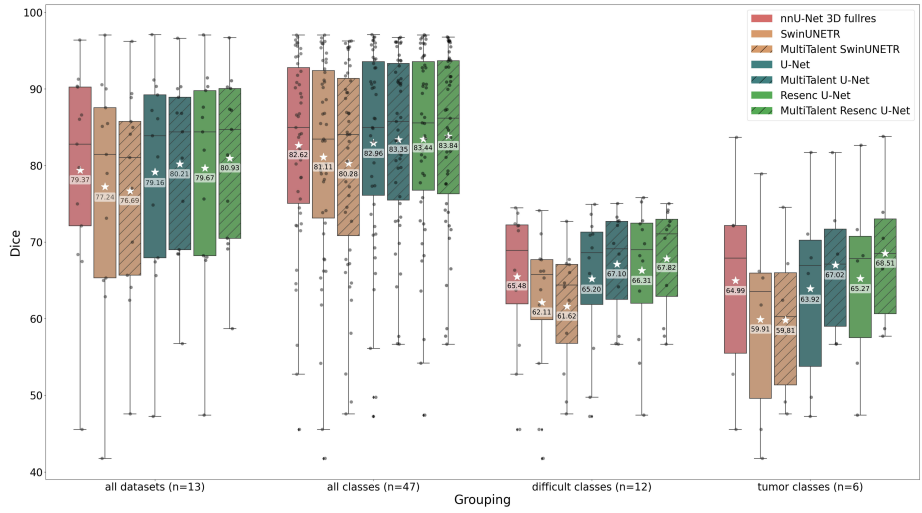}
\captionsetup{type=figure}
\captionof{figure}{Dice scores for all datasets, all classes, and classes of special interest. It should be noted that individual points within a boxplot corresponds to a different task. Difficult classes are those for which the default nnU-Net has a Dice below 75. The same color indicates the same architecture and the pattern implies training with multiple datasets using MultiTalent. The mean Dices are written on the Figure.  \label{fig:barplot_results}}

\end{figure}

\textbf{Multi-dataset training} results are presented in \autoref{fig:barplot_results}. In general, the convolutional architectures clearly outperform the transformer-inspired SwinUNETR. MultiTalent improves the performance of the purely convolutional architectures (U-Net and Resenc U-Net) and outperforms the corresponding baseline models that were trained on each dataset individually.  Since a simple average over all classes would introduce a biased perception due to the highly varying numbers of images and classes, we additionally report an average over all datasets. For example, dataset 7 consists of only 30 training images but has 13 classes, whereas dataset 6 has 126 training images but only 1 class. Table 4 in the appendix provides all results for all classes. Averaged over all datasets, the MultiTalent gains 1.26 Dice points for the Resenc U-Net architecture and 1.05 Dice points for the U-Net architecture. Compared to the default nnU-Net, configured without manual intervention for each dataset, the improvements are 1.56 and 0.84 Dice points. Additionally, in \autoref{fig:barplot_results} we analyzed two subgroups of classes. The first group includes all "difficult" classes for which the default nnU-Net has a Dice smaller than 75 (labeled by a "d" in Table 4 in the appendix). The second group includes all cancer classes because of their clinical relevance. Both class groups, but especially the cancer classes, experience notable performance improvements from MultiTalent. For the official BTCV test set in \autoref{BTCV_leaderboard}, MultiTalent outperforms all related work that have also incorporated multiple datasets during training,  proving that MultiTalent is substantially superior to related approaches. The advantages of MultiTalent include not only better segmentation results, but also considerable time savings for training and inference due to the simultaneous prediction of all classes. The training is 6.5 times faster and the inference is around 13 times faster than an ensemble of models trained on 13 datasets.\\ \\
\textbf{Transfer learning results} are found in \autoref{transfer_table}, which compares the fine-tuned 5-fold cross-validation results of different pre-training strategies for three different models on three datasets. The MultiTalent pre-training is highly beneficial for the convolutional models and outperforms all unsupervised baselines. Although MultiTalent was trained with a substantially lower amount of manually annotated structures (\raisebox{-0.9ex}{\~{}}3600 vs. \raisebox{-0.9ex}{\~{}} $10^5$ annotations), it also exceeds the supervised pre-training baseline. Especially for the small multi-organ dataset, which only has 30 training images (BTCV), and for the kidney tumor (KiTs19), the MultiTalent pre-training boosts the segmentation results. In general, the results show that supervised pre-training can be beneficial for the SwinUNETR as well, but pre-training on the large TotalSegmentator dataset works better than the MD pre-training. 
For the AMOS dataset, no pre-training scheme has a substantial impact on the performance. We suspect that it is a result of the dataset being saturated due to its large number of training cases. The Resenc U-Net pre-trained with MultiTalent, sets a new state-of-the-art on the BTCV leaderboard\footnote[1]{Assuming that no additional private data from the same data domain has been used.}(\autoref{BTCV_leaderboard}).

\begin{table}
\centering
\captionsetup{type=table}
\captionof{table}{Official BTCV test set leaderboard results, Dice and 95\% Hausdorff Distance. * indicates usage of multiple datasets. We submitted both a 5-fold cv ensemble and a single model to improve comparability to related methods.\label{BTCV_leaderboard}}
\begin{adjustbox}{width=0.75\textwidth}
\centering
\begin{tabular}{llcc}
\toprule
Method & \#models &Avg. Dice &  Avg. HD95 \\ \midrule
\multirow{2}{*}{nnU-Net (cascaded \& fullres) \cite{isensee_nnu-net_2021}} & 2 models, each  &\multirow{2}{*}{88.10} & \multirow{2}{*}{17.26} \\
& 5-fold ensemble & &\\
UNETR \cite{clip_medicalseg,hatamizadeh2021unetr}& single model &81.43  & -  \\
SwinUNETR \cite{clip_medicalseg,swinunetr} & single model &82.06 & - \\
Universal Model* \cite{clip_medicalseg} & single model &86.13 & - \\
DoDNet (pretrained*) \cite{zhang_dodnet_2020}&single model &86.44 & 15.62  \\ 
PaNN* \cite{zhou2019prioraware}  & single model&84.97 & 18.47  \\ 
\hdashline
MultiTalent Resenc U-Net* &single model &88.82 & 16.35  \\
MultiTalent Resenc U-Net* &5-fold ensemble &88.91 & \textbf{14.68} \\
\hdashline
Resenc U-Net (pre-trained MultiTalent*) & 5-fold ensemble &\textbf{89.07} &15.01 \\
\bottomrule
\end{tabular}
\end{adjustbox}
\end{table}

\begin{table}
\centering
\captionof{table}{5-fold cross validation results of different architectures and pretraining schemes. We used the original (org.) SwinUNETR implementation and the provided self-supervised weights as additional baseline \cite{swinunetr}. We applied a one sided paired t-test for each pretraining scheme compared to training from scratch. \label{transfer_table}}
\begin{adjustbox}{width=1\textwidth}
\centering
\begin{tabular}{ l l :c c: c c: c c: c c}
\toprule
 \multirow{2}{*}{Architecture}& \multirow{2}{*}{Pretraining scheme} & \multicolumn{2}{c:}{BTCV} & \multicolumn{2}{c:}{AMOS} & \multicolumn{4}{c}{KiTs19} \\ 
  &  & Dice avg.& p & Dice avg.& p &  Kidney Dice & p & Tumor Dice & p \\
\hline
SwinUNETR & from scratch & 74.27& & 86.04& &87.69 &  &46.56&\\
org. implement. & self-supervised \cite{swinunetr} &74.71& 0.30 & 86.11 &0.20& 87.62 & 0.55 &43.64& 0.97\\
\hdashline
\multirow{3}{*}{SwinUNETR} & from scratch & 81.44& & 87.59& &95.97 & &76.52&\\
 & supervised (\raisebox{-0.9ex}{\~{}}$10^5$ annot.) \cite{totalseg} & 83.08& $<0.01$&88.63& $<0.01$&96.36 &0.01 &80.30&$<0.01$ \\
 & \textbf{MultiTalent}(\raisebox{-0.9ex}{\~{}}3600 annot.)  &82.14& 0.02&87.32 &1.00 &96.08 & 0.28&76.56& 0.48\\
\hdashline
\multirow{4}{*}{U-Net} & from scratch& 83.76 && 89.40 && 96.56 && 80.69&\\
 & self-supervised \cite{ZHOU2021101840} & 84.01&0.11&  89.30&0.92& 96.59 &0.35& 80.91&0.39\\
 & supervised (\raisebox{-0.9ex}{\~{}}$10^5$ annot.) \cite{totalseg} & 84.22 &0.01 &89.66 & $<0.01$&96.72 &0.09 &82.48&0.02\\
 & \textbf{MultiTalent} (\raisebox{-0.9ex}{\~{}}3600 annot.) & 84.41& $<0.01$&89.60 &$<0.01$& 96.81 &0.04& 83.03&$<0.01$\\
\hdashline
\multirow{4}{*}{Resenc U-Net} & from scratch & 84.38 & &89.71& &96.83 && 83.22& \\
 & self-supervised \cite{ZHOU2021101840} & 84.27 &0.72& 89.70&0.64& 96.82 &0.56& 83.53 &0.35\\
 & supervised (\raisebox{-0.9ex}{\~{}} $10^5$ annot.) \cite{totalseg}&84.79 &0.04& \textbf{89.91} &$<0.01$& 96.85 &0.31& 83.73&0.23\\
 & \textbf{MultiTalent} (\raisebox{-0.9ex}{\~{}}3600 annot.) & \textbf{84.92} &0.03& 89.81 &0.16& \textbf{96.89} &0.04&\textbf{84.01}&0.12\\
\bottomrule
\end{tabular}
\end{adjustbox}

\end{table}

\section{Discussion}
MultiTalent demonstrates the remarkable potential of utilizing multiple publicly available partially labeled datasets to train a foundation medical segmentation network, that is highly beneficial for pre-training and finetuning various segmentation tasks.
MultiTalent surpasses state-of-the-art single-dataset models and outperforms related work for multi dataset training, while retaining conflicting annotation protocol properties from each dataset and allowing overlapping classes. Furthermore, MultiTalent takes less time for training and inference, saving resources compared to training many single dataset models. In the transfer learning setting, the feature representations learned by MultiTalent boost segmentation performance and set a new state-of-the-art on the BTCV leaderboard.
The nature of MultiTalent imposes no restrictions on additional datasets, which allows including any publicly available datasets (e.g. AMOS and TotalSegmentator). This paves the way towards holistic whole body segmentation model that is even capable of handling pathologies. 

\section*{Acknowledgements}
Part of this work was funded by Helmholtz Imaging (HI), a platform of the Helmholtz Incubator on Information and Data Science.
\bibliographystyle{splncs04}
\bibliography{bibfile.bib}

\newpage

\section*{Appendix}
\begin{table}
\centering
\captionsetup{type=table}
\captionof{table}{An overview of the 13 datasets that were used for the multi-class training. The last two datasets were only used for transfer learning experiments.} \label{tab_datasets}
\begin{adjustbox}{width=0.9\textwidth}
\begin{tabular}{l|l|c|c|c}
\toprule
 &   & Train &  & \\ 
Dataset & Labels & images & Median shape& Spacing [mm]\\
\hline
01 Decathlon Task 03 \cite{Antonelli2022}&   Liver, Liver tumor & 131  & 432x512x512&(1, 0.77, 0.77)\\ 
02 Decathlon Task 06 \cite{Antonelli2022}&   Lung nodules & 63 & 252x512x512 & (1.24, 0.79, 0.79)\\
03 Decathlon Task 07 \cite{Antonelli2022}&   Pancreas, Pancreas tumor & 281 & 93x512x512 & (2.5, 0.80, 0.80)\\
04 Decathlon Task 08 \cite{Antonelli2022} &   Hepatic vessels, Hepatic tumor & 303 & 49x512x512 & (5, 0.80, 0.80)\\
05 Decathlon Task 09 \cite{Antonelli2022}&   Spleen & 41 & 90x512x512 & (5, 0.79, 0.79) \\
06 Decathlon Task 10 \cite{Antonelli2022}&  Colon cancer & 126 & 95x512x512 &  (5, 0.78, 0.78)\\
07 BTCV \cite{BTCV} & 13 Abdominal Organs 
& 30& 128x512x512 & (3, 0.76, 0.76)\\
08 Pelvis \cite{BTCV} & Uterus, Bladder, Rectum, Small Bowel  & 30 & 180x512x512 & (2.5, 0.98, 0.98)\\
09 BTCV2 \cite{BTCV2} & 8 Abdominal Organs 
& 73 \tablefootnote[1]{Originally 90 images, but we removed duplicated images of the BTCV test set} & 185x512x512 & (3, 0.79, 0.79)\\
10 StructSeg Task 3 \cite{structseg} & 5 thoracic organs
& 50 & 95x512x512 & (5, 1.17, 1.17)\\
11 SegTHOR \cite{lambert2019segthor} & Heart, Aorta, Esophagus, Trachea & 40 & 178x512x512 & (2.5, 0.98, 0.98)\\
12 NIH-Pan \cite{roth2015deeporgan,clark_cancer_2013,NHpancreas} & Pancreas & 82& 217x512x512 & (1, 0.86, 0.86)\\
13 KiTS19 \cite{heller2020kits19} & Kidney, Kidney Tumor & 210 & 107x512x512 & (3, 0.78, 0.78)\\
Amos \cite{ji2022amos}& 15 abdominal organs & 300 & 104x512x512 & (5, 0.68, 0.68)\\
TotalSegmentator \cite{totalseg}& 104 classes & 1204 & 231x231x240& (1.5, 1.5, 1.5)\\ 

\bottomrule
\end{tabular}
\end{adjustbox}

\vspace*{0.5 cm}

\centering
\captionsetup{type=figure}
\includegraphics[width=0.85\textwidth]{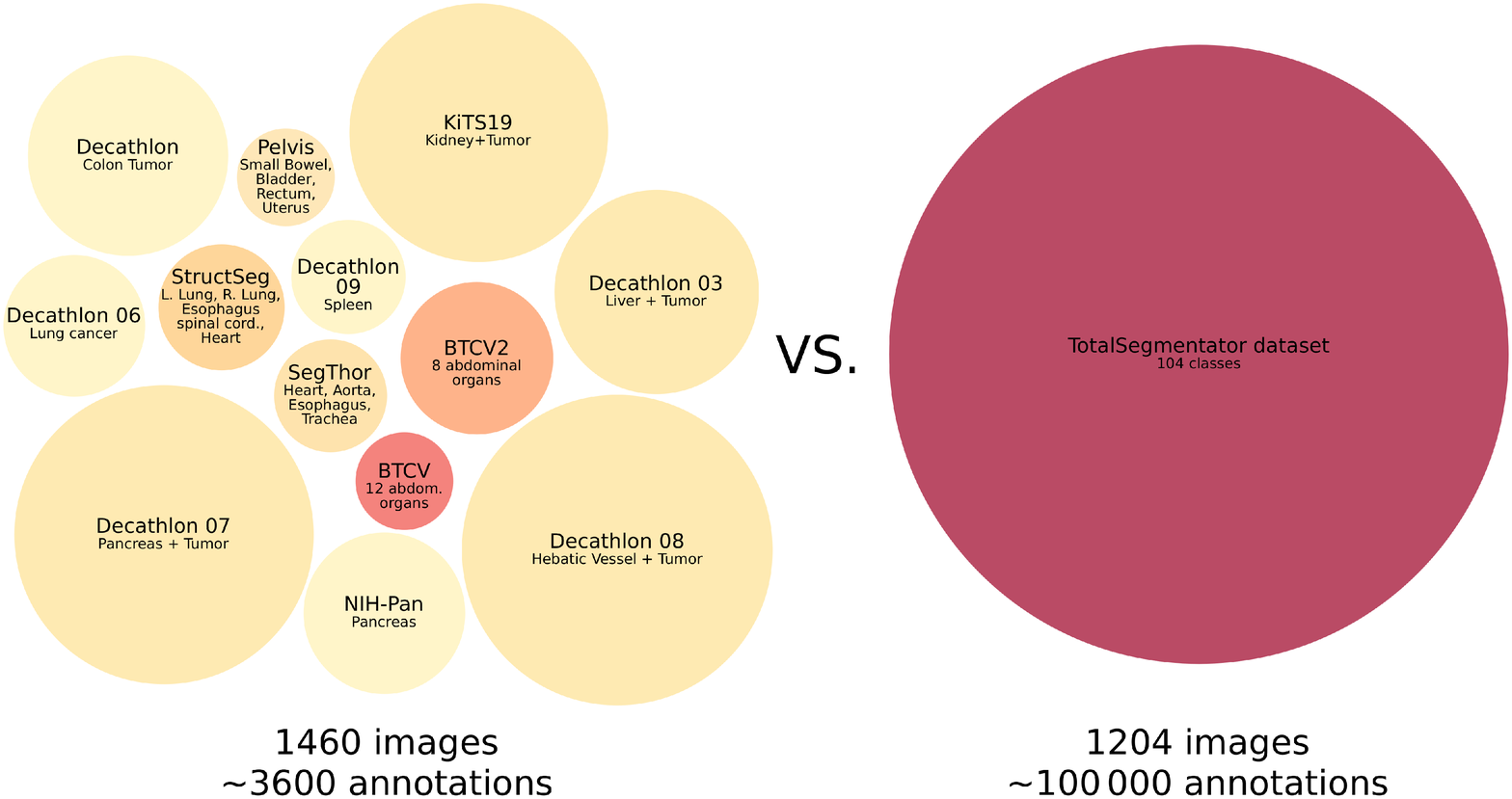}
\captionof{figure}{The circle area corresponds to each dataset size and the color indicates the number of annotated classes. MultiTalent was trained on 1460 images with about 3600 annotations. Whereas, the TotalSegmentator dataset consists of 1204 images with about $10^5$ annotations \cite{totalseg}.}
\label{fig:dataset_overview}

\end{table}

\include{bigtable.tex} 
\end{document}

%% file: bigtable.tex
\begin{table}
\centering
\caption{Dice (5 fold) for all classes within all datasets. All networks but the default nnUNet are trained with batch size 4 and patch size [96,192,192]. MD indicates the MultiTalent networks that were trained with multiple datasets together. We indicated all cancer (c) and difficult (d) classes, for which the default nnU-Net has a Dice below 75. We added the parameter and flop count for each network (for D1 as an example of the per dataset models).}
\label{big_table}
\adjustbox{width = 0.81\textwidth}{
\begin{tabularx}{\textwidth}{ l| Y|  X X X X X X p{1cm} }

\toprule
Classes & Class groups & nnU-Net &  SwinU &  U-Net &  Resenc &  MD SwinU &  MD U-Net &  MD Resenc \\
\hdashline
\multicolumn{1}{r|}{Parameters $* 10^6$}& &31.20 &62.19  &29.29 &69.30 &62.19 & 29.32&69.34\\
\multicolumn{1}{r|}{Flops $* 10^{12}$}&&0.81&1.28&0.72&1.19&1.29&0.72&1.20\\
\hline
1 liver w/o cancer      &  - &   95.71 &  96.14 & 96.34 &   \textbf{96.48} &     95.94 &    96.10 &      96.27 \\
1 liver tumor           & d, c &   63.72 &  61.78 & 65.89 &   \textbf{67.53} &     58.09 &    65.83 &      66.50 \\
2 lung\_nodule            & d, c &   \textbf{72.11} &  65.31 & 67.94 &   68.22 &     62.40 &    68.43 &      70.49 \\
2 pancreas w/o cancer     &-  &   \textbf{82.17} &  80.19 & 81.52 &   81.91 &     79.04 &    81.35 &      81.85 \\
3 pancreas cancer         & d, c&   52.74 &  45.53 & 49.73 &   54.18 &     49.12 &    56.67 &      \textbf{57.71} \\
4 hepatic vessel          & d&   \textbf{64.56} &  63.77 & 63.78 &   63.60 &     64.14 &    64.15 &      64.33 \\
4 liver cancer            & d, c&   72.17 &  66.18 & 71.04 &   71.62 &     67.18 &    72.78 &      \textbf{73.88} \\
5 spleen                  &- &   96.38 &  97.02 & \textbf{97.08} &   97.05 &     96.20 &    96.61 &      96.67 \\
6 colon cancer            &d, c &   45.53 &  41.75 & 47.23 &   47.40 &     47.56 &    56.73 &      \textbf{58.69} \\
7 spleen                  & -&   90.83 &  91.14 & 92.72 &   92.34 &     91.04 &    \textbf{93.75} &      93.61 \\
7 right kidney            &- &   89.39 &  88.59 & \textbf{90.93} &   90.33 &     86.93 &    90.90 &      90.89 \\
7 left kidney             & -&   86.75 &  86.80 & 90.18 &   90.65 &     86.83 &    \textbf{90.75} &      90.47 \\
7 gallbladder             &d &   66.32 &  66.21 & 69.24 &   69.78 &     67.73 &    69.82 &      \textbf{72.12} \\
7 esophagus               &- &   78.40 &  77.87 & 78.56 &   78.96 &     77.87 &    \textbf{79.73} &      79.28 \\
7 liver            & - &   95.57 &  95.36 & 95.74 &   95.95 &     95.14 &    96.12 &   \textbf{96.23} \\
7 stomach                 &- &  88.16 &  86.71 & 90.87 &   92.76 &     86.77 &    91.11 &         \textbf{92.83} \\
7 aorta                   &- &   92.29 &  92.15 & \textbf{92.75} &   \textbf{92.75} &     89.92 &    91.32 &      91.60  \\
7 vena cava inferior         & - &   86.38 &  85.35 & 86.16 &   86.73 &     84.34 &    86.26 &   \textbf{87.31} \\
7 portal \& splenic vein    & - &    76.59 &  73.84 & 77.29 &   77.74 &     73.88 &    \textbf{77.76} &      77.58  \\
7 pancreas & -&  81.76 &  79.26 & 83.10 &   83.87 &     82.27 &    84.62 & \textbf{84.92} \\
7 right adrenal gland      & d &  71.48 &  67.69 & 70.90 &   72.37 &     66.02 &     \textbf{72.70} &      \textbf{72.70} \\
7 left adrenal gland       & d&   72.38 &  67.78 & 72.03 &   \textbf{72.77} &     64.65 &    72.17 &      71.62 \\
8 bladder                & - &   88.93 &  88.26 & 89.26 &   89.41 &     84.02 &    90.73 &      \textbf{91.59} \\
8 uterus                 & - &   \textbf{80.72} &  78.94 & 80.34 &   80.57 &     76.06 &    79.64 &      78.89 \\
8 rectum                 & d &   73.77 &  71.07 & 73.62 &   \textbf{75.24} &     67.06 &    73.25 &      74.17 \\
8 small bowel           &  d &    56.52 &  54.16 & 56.12 &   57.27 &     52.77 &    \textbf{57.68} &      56.67 \\
9 spleen                  &- &   94.59 &  93.44 & 95.47 &   \textbf{95.86} &     93.82 &    95.53 &      95.49 \\
9 left kidney            & - &   93.29 &  92.60 & 93.81 &   \textbf{94.38} &     91.69 &    93.71 &      93.72 \\
9 gallbladder             & -&   78.44 &  78.17 & 81.30 &   81.07 &     77.21 &    80.63 &      \textbf{81.76} \\
9 esophagus               & d &  74.46 &  74.10 & 74.92 &   \textbf{75.79} &     72.71 &    75.03 &      75.01 \\
9 liver                   &- &   96.03 &  95.90 & 96.32 &   \textbf{96.50} &     95.41 &    96.09 &      96.19 \\
9 stomach                 &- &   91.77 &  90.90 & \textbf{93.32} &   93.09 &     88.75 &    92.17 &      92.84 \\
9 pancreas                &- &   84.04 &  83.20 & 84.95 &   85.48 &     84.05 &    85.74 &      \textbf{86.16} \\
9 duodenum                &- &   75.62 &  72.47 & 77.35 &   \textbf{78.70} &     68.96 &    75.92 &      77.60 \\
10 left lung              & - &   95.99 &  95.89 & 95.99 &   95.97 &     95.48 &    95.95 &       \textbf{96.03} \\
10 right lung             & - &   96.74 &  96.65 & 96.77 &   96.74 &     96.25 &    96.73 &      \textbf{96.78} \\
10 heart                   & -&   \textbf{94.17} &  93.87 & 93.91 &   94.05 &     93.02 &    93.66 &      94.01 \\
10 esophagus               &- &   80.19 &  79.97 & 81.30 &   \textbf{81.39} &     78.91 &    80.87 &      81.06 \\
10 bronchies               &- &   \textbf{84.16} &  83.44 & 83.88 &   83.51 &     83.13 &    83.23 &      83.64 \\
10 spinaL cord nerve  &- & 90.14 &  \textbf{90.17} & 90.08 &   90.14 &     89.51 &    89.69 &      89.94 \\
11 esophagus              & - &     84.95 &  83.18 & 84.83 &   \textbf{85.56} &     80.53 &    83.21 &      84.61 \\
11 heart                  & - &   95.27 &  94.67 & 95.11 &   \textbf{95.29} &     93.36 &    94.40 &      94.83 \\
11 trachea                & - &   90.55 &  90.61 & 90.61 &   90.59 &     89.50 &    90.50 &      \textbf{90.96} \\
11 aorta                  & - &   94.26 &  93.70 & 94.11 &   \textbf{94.29} &     92.07 &    93.00 &      93.82 \\
12 pancreas               & - &   86.59 &  85.47 & 86.10 &   86.27 &     85.74 &    86.80 &      \textbf{87.21} \\
13 kidneys w/o tumor   &- &    \textbf{97.02} &  96.19 & 96.72 &   96.92 &     95.37 &    96.11 &      96.34 \\
13 kidney tumor            &c &    83.67 &  78.90 & 81.70 &   82.64 &     74.54 &    81.68 &      \textbf{83.79} \\
\hline
Class average                     & -&  82.62 &  81.11 & 82.96 &   83.44 &     80.28 &    83.35 &      \textbf{83.84} \\
\bottomrule
\end{tabularx}}

\end{table}